\documentstyle[amssymb,aps]{revtex}

\begin{document}
\title{Pair production with electron capture in peripheral collisions of
relativistic heavy ions }
\author{C.A. Bertulani$^{(a,b,\ast )}$ and D. Dolci$^{(a)}$}
\address{$^{a}$Instituto de F\'{i}sica,Universidade Federal do Rio de Janeiro, \\
21945-970 Rio de Janeiro, RJ, Brazil}
\address{$^{b}$ Brookhaven National Laboratory, Upton, NY, USA }
\date{\today }
\maketitle

\begin{abstract}
The production of electron-positron pairs with the capture of the electron
in an atomic orbital is investigated for the conditions of the Relativistic
Heavy Ion Collider (RHIC) and the Large Hadron Collider LHC). Dirac wave
functions for the leptons are used, taking corrections to orders of $Z\alpha 
$ into account. The dependence on the transverse momentum transfer is
studied and the accuracy of the equivalent photon approximation is discussed
as a function of the nuclear charge.
\end{abstract}

\bigskip


\bigskip


\section{Introduction}

Peripheral collisions with relativistic heavy ions are a source of several
interesting physical processes and have attracted a great number of
theoretical and experimental work (see, e.g., \cite{BB94}, and references
therein). One of the processes of interest is pair production by the strong
electromagnetic field of large Z nuclei in relativistic heavy ion colliders.
In the case of production of free pairs the theoretical interest resides in
the non-perturbative character of the process. Different techniques have
been used to study this problem theoretically, but we will not discuss them
here. Another case of general interest is the production of pairs in which
the electron is captured in a bound state of one of the colliding nuclei 
\cite{Be88} (see also ref. \cite{BHT98} and references therein). The
interest here arises from its consequence for luminosity changing of the
relativistic ion beams in colliders like RHIC (Relativistic Heavy Ion
Collider) at Brookhaven, and the LHC (Large Hadron Collider) at CERN (for a
recent analysis, see \cite{Sp00}). A more striking application of this
process was the recent production of antihydrogen atoms using relativistic
antiproton beams \cite{Ba96}. In that case the positron is produced and
captured in an orbit of an antiproton.

Early calculations of pair production with capture used perturbation theory 
\cite{AB87,Be88}, the only difference being the way the distortion effects
for the positron wavefunction were taken into account. Evidently the
calculation is best performed using the frame of reference of the nucleus
where the electron is captured. Many other calculations have been performed 
\cite{Bec87,RB89,Mo91,BRW93,Ru93,As94,Mo95,Ag97,Ba97}. Some of them used
non-perturbative approaches, e.g., coupled-channels calculations. However,
in contrast to the production of free pairs, pair-production with atomic
capture of the electron is well described in perturbation theory. The large
coupling constant for large nuclei ($Z\alpha \sim 1$) does not matter here,
since the scaled matrix element divided by $Z\alpha $, i.e. $M/(Z\alpha )$,
is much smaller in this case than it is for the production of free pairs. In
other words, the overlap between the positron and the electron wavefunctions
is much less than that for the production of free pairs. The first terms of
the perturbation series are small enough to neglect the inclusion of higher
order terms \cite{Be88}. Indeed it was explicitly shown in ref.  \cite{Ba97}
that non-perturbative calculations modify the cross sections at most by 1\%. 

The theoretical difficulty here is to properly account for the distortion of
the positron wavefunction in the field of the nucleus where the electron is
captured. In this article we study this property more closely, and compare
our results with the calculations of ref. \cite{Be88}, which were based on
the Sommerfeld-Maue wavefunctions for the positron \cite{Be54}. We also
investigate the accuracy and limitations of the equivalent photon
approximation (EPA), what will gives us a clear understanding of the effects
of distortion. We have closely used the technique of ref. \cite{BeB98} for
the study of anti-hydrogen production cross sections at LEAR/CERN. We show
that a commonly used approximation obtained in ref. \cite{Be88} yields
results up to a factor 5 too small for large ion charges.

In section 2 we deduce the cross sections for pair production with K-shell
capture and obtain the cross sections in terms of longitudinal and
transverse components. In section 3 we compare the calculation with that of
pair production with capture by real photons. This allows us to disclose the
role of distortion effects. We also show that the transverse part of the
virtual photon cross section is much larger than the longitudinal one. In
section 4 we calculate total cross sections and compare them to the results
obtained in ref. \cite{Be88}, to the equivalent photon approximation and to
recent experimental data .  Our conclusions are given in section 5.

\section{Pair Production with K-shell capture}

In the semiclassical approach we assume that the relativistic ions move
along straight lines. Although the laboratory system is a more symmetric
frame of reference, the calculation is best accomplished in the frame of
reference of one of the ions - the one in which the electron is captured.
The reason is that the effects of distortion of the positron wavefucntion
are very important, and cannot be neglected. This distortion effect is most
naturally taken into account in the rest frame of the nucleus.

The relationship between the Lorentz contraction factor associated with the
relative velocity between the colliding nuclei, $\gamma $, and the collider
energy per nucleon in GeV , $E/A$, is given by $\gamma =2(1.0735\times
E/A)^{2}$, for $E/A\gg 1.$ For example, at the RHIC, $E/A=100$, and \ $%
\gamma =2.3\times 10^{4}$, while at the LHC, $E/A=3\times 10^{3}$, and \ $%
\gamma \cong 2.07\times 10^{7}$. According to refs.\cite{Be88,BeB98} the
probability amplitude for pair production with capture for a collision with
impact parameter ${\bf b}$ is given by

\begin{equation}
a_{1st}=\frac{Ze}{i\pi }\int d^{2}q_{t}\;\frac{e^{i{\bf q}_{t}{\bf .b}}}{%
q_{t}^{2}+\left( \omega /{\rm v}\gamma \right) ^{2}}\;F({\bf Q})\;,
\label{amp1}
\end{equation}
where ${\bf q}_{t}$ is a \ transverse momentum vector with two components, $%
{\bf Q}\equiv ({\bf q}_{t},\omega /{\rm v})$ is a 3-component momentum
vector with the third component equal to $\omega /{\rm v},\;$and $\omega $
is the total energy used in the pair production. It is important to keep the
relative velocity between the colliding nuclei, ${\rm v}$, in the right
places for the moment (although ${\rm v}\cong c=1$) , since important
combinations of 1 and ${\rm v}$ will lead to Lorentz \ factors $\gamma
=\left( 1-{\rm v}^{2}\right) ^{-1/2}$ in subsequent steps of the
calculation. In the case of K-shell capture $\omega =\varepsilon +m-I_{K}$, $%
I_{K}$ is the ionization energy of the K-shell electron, and $\varepsilon $
is the positron total energy $\varepsilon =\sqrt{p^{2}+1}$ (from now on we
will use natural units, so that $\hbar =c=m=1$). The form factor $F({\bf Q})$%
\ \ is given in terms of the lepton transition current $j_{\mu }(r)=e%
\overline{\Psi ^{(+)}}\gamma _{\mu }\Psi ^{(-)}$ as

\begin{equation}
F({\bf Q})=ie\;K_{\mu }\int d^{3}r\;\overline{\Psi ^{(+)}}({\bf r})\;e^{i%
{\bf Q.r}}\;\gamma _{\mu }{\bf \;}\Psi ^{(-)}({\bf r})\;,  \label{FQ}
\end{equation}
where $K_{\mu }\equiv \left( 0,{\bf K}\right) $, with ${\bf K}=\left( {\bf q}%
_{t}/\omega ,1/{\rm v}\gamma ^{2}\right) $. For details of derivation of
eqs. (\ref{amp1}-\ref{FQ}), see refs. \cite{Be88,BeB98}. The cross section
is obtained by integrating the square of the expression (\ref{amp1}) over
all possible impact parameters:

\begin{equation}
\sigma =\sum_{spins}\int \;\left| a_{1st}\right|
^{2}d^{2}b=4Z^{2}e^{2}\sum_{spins}\int d^{2}q_{t}\;\frac{\left| F({\bf Q}%
)\right| ^{2}}{({\bf Q}^{2}-\omega ^{2})^{2}}\;.  \label{sigma1}
\end{equation}

For the positron wavefunction we use a plane wave and a correction term to
account for the distortion due to the nucleus charge. The correction term is
considered to be proportional to $Ze^{2}$. The wavefunction is then given by

\begin{equation}
\Psi ^{(+)}=N_{+}\left[ {\rm v\ }e^{i{\bf p.r}}+\Psi ^{\prime }\right] \;,
\label{phip}
\end{equation}
where\ $N_{+}$ \ is a normalization constant.

Only the Fourier transform of $\Psi ^{\prime }$ will enter the calculation.
This Fourier transform can be deduced directly from the Dirac equation for
the positron in the presence of a Coulomb field of a nucleus:

\begin{equation}
\left( \gamma ^{0}\varepsilon +i\overrightarrow{\gamma}{.}\overrightarrow{%
\nabla }{\bf +}m\right) \Psi ^{\prime }=-\frac{Ze^{2}}{r}\;\gamma ^{0}{\rm v}%
\;e^{i{\bf p.r}},
\end{equation}
Applying on both sides of this equation the operator $\left( \gamma
^{0}\varepsilon +i\overrightarrow{\gamma }{\bf .}\overrightarrow{\nabla }%
{\bf -}m\right) $ we get

\begin{equation}
\left( \Delta +{\bf p}^{2}\right) \Psi ^{\prime }=-Ze^{2}\left( \gamma
^{0}\varepsilon +i\overrightarrow{\gamma }{\bf .}\overrightarrow{\nabla }%
{\bf -}m\right) (\gamma ^{0}{\rm v})\frac{e^{i{\bf p.r}}}{r}\;.
\end{equation}
Multiplying by $e^{-i{\bf q.r}}\;$and integrating over $d^{3}r$ we get

\begin{equation}
\left( {\bf p}^{2}-{\bf q}^{2}\right) \Psi _{{\bf q}}^{\prime }=-Ze^{2}\left[
2\gamma ^{0}\varepsilon -\overrightarrow{\gamma }{\bf .(q-p)}\right] (\gamma
^{0}{\rm v})\frac{4\pi }{({\bf q-p})^{2}}\;,
\end{equation}
where we have used the identity $\gamma ^{0}\left( \gamma ^{0}\varepsilon -i%
\overrightarrow{\gamma }{\bf .}\overrightarrow{\nabla }{\bf +}m\right)
(\gamma ^{0}{\rm v})=0.$ Thus,

\begin{equation}
\overline{\Psi _{{\bf q}}^{\prime }}\equiv \left( \Psi _{{\bf q}}^{\prime
}\right) ^{\dagger }\gamma ^{0}=4\pi Ze^{2}\;\overline{{\rm v}}\;\frac{%
2\gamma ^{0}\varepsilon +\overrightarrow{\gamma }{\bf .(q-p)}}{({\bf q-p}%
)^{2}({\bf q}^{2}-{\bf p}^{2})}\gamma ^{0}\;.  \label{phipq}
\end{equation}
We will need this result later to calculate $F({\bf Q})$. The normalization
constant $N_{+},$\ which accounts for the distortion of the wavefunction to
all orders is given by \cite{Be88}

\begin{equation}
N_{+}=\exp \left[ \pi a_{+}/2\right] \Gamma \left( 1+ia_{+}\right)
\;,\;\;\;\;\;\;\;\;a_{+}=Ze^{2}/{\rm v}_{+}\;,  \label{NN}
\end{equation}
where ${\rm v}_{+}$ \ is the positron velocity in the frame of reference of
the nucleus where the electron is capture.

For the electron we use the distorted wavefunction described in ref. \cite
{Be79}, i.e.,

\begin{equation}
\Psi ^{(-)}=\left[ 1-\frac{i}{2}\;\gamma ^{0}\overrightarrow{{\bf \gamma }}%
{\bf .}\overrightarrow{{\bf \nabla }}\right] {\rm u}\Psi
_{non-r}\;,\;\;\;\;\;\;{\rm \;\;where\;\;}\;\;\;\;\Psi _{non-r}=\frac{1}{%
\sqrt{\pi }}\left( \frac{Z}{a_{H}}\right) ^{3/2}\exp \left( -Zr/a_{H}\right)
\;,  \label{phie}
\end{equation}
with $a_{H}=1/e^{2}$ being the hydrogen Bohr radius.

Using the positron and the electron wavefunctions as given by eqs. (\ref
{phip}) and (\ref{phie}), we get

\begin{eqnarray}
F({\bf Q}) &=&ieN_{+}\int d^{3}r\;\left\{ \overline{{\rm v}}\;e^{-i{\bf p.r}%
}+\overline{\Psi ^{\prime }}\right\} \left( \gamma _{\mu }K^{\mu }\right)
\;e^{i{\bf Q.r}}\;  \nonumber \\
\times &&\left\{ 1-\frac{i}{2m}\;\gamma ^{0}\overrightarrow{\gamma }{\bf .}%
\overrightarrow{\nabla }\right\} \;{\rm u}\Psi _{non-r}({\bf r})  \nonumber
\\
&\simeq &ieN_{+}\int d^{3}r\;\Bigg\{\overline{{\rm v}}\;e^{i({\bf Q}-{\bf %
p).r}}\left( \gamma _{\mu }K^{\mu }\right) \;\left[ \left( 1-\frac{i}{2m}%
\;\gamma ^{0}\overrightarrow{\gamma }{\bf .}\overrightarrow{\nabla }\right) %
\right] {\rm u}\Psi _{non-r}({\bf r})\;  \nonumber \\
+ &&\overline{\Psi ^{\prime }}\;e^{i{\bf Q.r}}\;\left( \gamma _{\mu }K^{\mu
}\right) {\rm u}\Psi _{non-r}({\bf r})\Bigg\}\;,
\end{eqnarray}
where in the last equation we neglected terms of highest orders in $Z\alpha
. $

Integrating by parts yields

\begin{equation}
F({\bf Q})\simeq ieN_{+}\Bigg\{\overline{{\rm v}}\;\left( \gamma _{\mu
}K^{\mu }\right) \;\left[ \left( 1+\frac{1}{2m}\;\gamma ^{0}\overrightarrow{%
\gamma }{\bf .(Q-p)}\right) {\rm u\;}\left( \Psi _{non-r}\right) _{{\bf Q-p}}%
\right] \;+\overline{\Psi _{{\bf Q}}^{\prime }}\left( \gamma _{\mu }K^{\mu
}\right) {\rm u}\frac{1}{\sqrt{\pi }}\left( \frac{Z}{a_{H}}\right) ^{3/2}%
\Bigg\}\;.  \label{FQi}
\end{equation}

Since the corrections are to first order in $Z\alpha $, we have replaced $%
\Psi _{non-r}({\bf r})$ by its constant value $\left( Z/a_{H}\right)
^{3/2}\left( 1/\sqrt{\pi }\right) $ in the last term of eq. (\ref{FQi}). If
we would do the same in the first term it would be identically zero (except
for ${\bf Q=p}$). We need to keep the corrections in the wavefunctions at
least to first order in $Z\alpha $; for $\varepsilon \gg 1$ these
corrections yield a term of the same order as the plane-wave to the total
cross section, as we shall see later.

Inserting these results in eq. \ref{sigma1}, and using $\left[ \exp \left(
-r/a\right) \right] _{{\bf Q-p}}=8\pi /a\left( {\bf Q-p}\right) ^{4}$,\ \ we
get,

\begin{equation}
F({\bf Q})=ieN_{+}\;8\sqrt{\pi }\left( \frac{Z}{a_{H}}\right) ^{5/2}\;\frac{%
{\rm v}A{\rm u}}{\left( {\bf Q}-{\bf p}\right) ^{2}},  \label{FQ2}
\end{equation}
where

\begin{equation}
A=a(\gamma _{\mu }K^{\mu })+(\gamma _{\mu }K^{\mu })\gamma ^{0}\left( 
\overrightarrow{\gamma }{\bf .b}\right) +(\overrightarrow{\gamma }{\bf .d}%
)\gamma ^{0}(\gamma _{\mu }K^{\mu })\;,
\end{equation}
with

\begin{eqnarray}
a &=&\frac{1}{\left( {\bf Q-p}\right) ^{2}}-\frac{\varepsilon }{{\bf Q}^{2}-%
{\bf p}^{2}}\;,\;\;\;\;{\bf b}=\frac{{\bf Q-p}}{2\left( {\bf Q-p}\right) ^{2}%
}\;,\;\;\;\;{\rm and}  \nonumber \\
{\bf d} &=&-\frac{{\bf Q-p}}{2({\bf Q}^{2}-{\bf p}^{2})}\;.  \label{abd}
\end{eqnarray}
Inserting (\ref{FQ2}) into (\ref{sigma1}), with $\left| N_{+}\right|
^{2}=2\pi a_{+}/\left[ \exp \left( 2\pi a_{+}\right) -1\right] $, performing
the sum over spins with standard trace techniques, and using $%
p^{2}dp=\varepsilon pd\varepsilon $ ($p\equiv \left| {\bf p}\right| $), we
get

\begin{eqnarray}
\frac{d\sigma }{d\varepsilon d\Omega } &=&\frac{64}{\pi a_{H}^{5}{\rm v}_{+}}%
\;\frac{Z^{8}e^{6}p}{\left[ \exp \left( 2\pi Ze^{2}/{\rm v}_{+}\right) -1%
\right] }\int \frac{d^{2}q_{t}}{({\bf Q}^{2}-\omega ^{2})^{2}}\;\frac{1}{%
\left( {\bf Q-p}\right) ^{4}}  \nonumber \\
&\times &\Bigg\{\left( \varepsilon -1\right) \left[ \left( {\bf b-d}\right)
^{2}K^{2}+4\left( {\bf K.b}\right) \left( {\bf K.d}\right) \right]  \nonumber
\\
+ &&\left( \varepsilon +1\right) a^{2}K^{2}-2a\left[ 2\left( {\bf p.K}%
\right) \left( {\bf b.K}\right) +{\bf p}.({\bf b-d})K^{2}\right] \Bigg\}\;.
\end{eqnarray}
In this relation we have used $\varepsilon _{-}\sim 1$, neglecting the
electronic binding energy in the numerator of the expression inside the
integral.

Using the definition of $K_{\mu }$, it is illustrative to separate the
integrand of the above equation in terms of longitudinal and transverse
components:

\begin{equation}
\frac{d\sigma }{d\varepsilon d\Omega }=\frac{1}{\pi \omega }%
\;Z^{2}e^{2}\int_{0}^{\infty }d(q_{t}^{2})\;\frac{q_{t}^{2}}{({\bf Q}%
^{2}-\omega ^{2})^{2}}\;\Bigg[\frac{d\sigma _{\gamma ^{\ast }T}}{d\Omega }%
\left( {\bf Q},\omega \right) +\frac{d\sigma _{\gamma ^{\ast }L}}{d\Omega }%
\left( {\bf Q},\omega \right) +\frac{d\sigma _{\gamma ^{\ast }LT}}{d\Omega }%
\left( {\bf Q},\omega \right) \Bigg]\;,  \label{dsdedO}
\end{equation}

where

\begin{eqnarray}
\frac{d\sigma _{\gamma ^{\ast }T}}{d\Omega }\left( {\bf Q},\omega \right) &=&%
\frac{64\pi p}{a_{H}^{5}\omega {\rm v}_{+}}\frac{Z^{6}e^{4}}{\left[ \exp
\left( 2\pi Ze^{2}/{\rm v}_{+}\right) -1\right] }\frac{1}{\left( {\bf Q-p}%
\right) ^{4}}\;\Bigg\{\left( \varepsilon -1\right) \left[ \left( {\bf b-d}%
\right) ^{2}-\frac{\left( q_{t}-{\bf p.}\widehat{{\bf e}}_{t}\right) ^{2}}{%
\left( {\bf Q-p}\right) ^{2}\left( {\bf Q}^{2}-{\bf p}^{2}\right) }\right]
\label{st} \\
+ &&\left( \varepsilon +1\right) a^{2}+2a\left[ \frac{\left( q_{t}-{\bf p.}%
\widehat{{\bf e}}_{t}\right) {\bf p.}\widehat{{\bf e}}_{t}}{\left( {\bf Q-p}%
\right) ^{2}}-{\bf p}.({\bf b-d})\right] \Bigg\}\;,  \nonumber
\end{eqnarray}

\begin{eqnarray}
\frac{d\sigma _{\gamma ^{\ast }L}}{d\Omega }\left( {\bf Q},\omega \right) &=&%
\frac{64\pi p\omega }{a_{H}^{5}q_{t}^{2}\gamma ^{4}{\rm v}_{+}}\frac{%
Z^{6}e^{4}}{\left[ \exp \left( 2\pi Ze^{2}/{\rm v}_{+}\right) -1\right] }%
\frac{1}{\left( {\bf Q-p}\right) ^{4}}\;\Bigg\{\left( \varepsilon -1\right) %
\left[ \left( {\bf b-d}\right) ^{2}-\frac{\left( \omega /{\rm v}%
-p_{z}\right) ^{2}}{\left( {\bf Q-p}\right) ^{2}\left( {\bf Q}^{2}-{\bf p}%
^{2}\right) }\right]  \nonumber \\
+ &&\left( \varepsilon +1\right) a^{2}+2a\left[ \frac{p_{z}\left( \omega /%
{\rm v}-p_{z}\right) }{\left( {\bf Q-p}\right) ^{2}}-{\bf p}.({\bf b-d})%
\right] \Bigg\}\;,  \label{sl}
\end{eqnarray}

\begin{eqnarray}
\frac{d\sigma _{\gamma ^{\ast }LT}}{d\Omega }\left( {\bf Q},\omega \right)
&=&\frac{128\pi p}{a_{H}^{5}q_{t}\gamma ^{2}{\rm v}_{+}}\frac{Z^{6}e^{4}}{%
\left[ \exp \left( 2\pi Ze^{2}/{\rm v}_{+}\right) -1\right] }\frac{\left(
\omega /{\rm v}-p_{z}\right) }{\left( {\bf Q-p}\right) ^{6}}  \nonumber \\
&&\times \left\{ a\left[ \left( {\bf p.}\widehat{{\bf e}}_{t}\right) +p_{z}%
\frac{\left( q_{t}-{\bf p.}\widehat{{\bf e}}_{t}\right) }{\left( \omega /%
{\rm v}-p_{z}\right) }\right] -\left( \varepsilon -1\right) \frac{\left(
q_{t}-{\bf p.}\widehat{{\bf e}}_{t}\right) }{\left( {\bf Q}^{2}-{\bf p}%
^{2}\right) }\right\} \;,  \label{slt}
\end{eqnarray}

In these equations $\widehat{{\bf e}}=\widehat{{\bf q}}_{t}/q_{t}$ is a unit
vector in the transverse direction. The cross sections $\sigma _{\gamma
^{\ast }T}$, $\sigma _{\gamma ^{\ast }L}$, and $\sigma _{\gamma ^{\ast }LT}$
are interpreted as the cross sections for pair production with capture by
virtual transverse and longitudinal photons and a interference term,
respectively. Note that the transverse and longitudinal directions are {\it %
with respect to the beam axis}, not with respect to the photon momentum, as
is usually meant by this term. Only for $\gamma \gg 1$, this definition
agrees with the definition of transverse and longitudinal virtual photons.
However, this separation is very useful, as we will see next.

Now we turn to the relation to pair-production by real photons and the
relation to the equivalent photon approximation.

\section{Pair production with capture by real photons: role of distortion
effects}

The cross section for pair production with capture real photons is given by 
\cite{Be79}

\begin{equation}
d\sigma _{\gamma }=2\pi \left| V_{fi}\right| ^{2}\delta (\omega -\varepsilon
-1)\frac{d^{3}p}{\left( 2\pi \right) ^{3}}\;,  \label{sgr}
\end{equation}
where

\begin{equation}
V_{fi}=-e\sqrt{\frac{4\pi }{\omega }}%
\displaystyle\int %
d^{3}r\;\overline{\Psi ^{(+)}}({\bf r})\;e^{i\overrightarrow{{\bf \kappa }}%
{\bf .r}}\;\left( \widehat{{\bf e}}.\overrightarrow{\gamma }\right) \Psi
^{(-)}({\bf r})\;,  \label{Vfi}
\end{equation}
with $\overrightarrow{{\bf \kappa }}=\widehat{{\bf z}}\;\omega $, where $%
\widehat{{\bf z}}$ is the unit vector along the photon incident direction,
and $\widehat{{\bf e}}$ is photon polarization unit vector.

Using the positron and the electron wavefunctions from the previous section,
and performing similar steps, we get

\begin{equation}
V_{fi}=-eZ^{5/2}\frac{8\pi \sqrt{2}}{a_{H}^{5/2}\omega ^{1/2}}N_{+}\frac{%
{\rm v}Bu}{\left( \overrightarrow{{\bf \kappa }}{\bf -p}\right) ^{2}}\;,
\end{equation}
where

\begin{equation}
B=a_{\kappa }(\overrightarrow{\gamma }{\bf .}\overrightarrow{{\bf \kappa }}%
)+(\overrightarrow{\gamma }{\bf .}\overrightarrow{{\bf \kappa }})\gamma
^{0}\left( \overrightarrow{\gamma }{\bf .b}_{\kappa }\right) +(%
\overrightarrow{\gamma }{\bf .d}_{\kappa })\gamma ^{0}(\overrightarrow{%
\gamma }{\bf .}\overrightarrow{{\bf \kappa }})\;.
\end{equation}
where $a_{\kappa }$, ${\bf b}_{\kappa }$, and ${\bf d}_{\kappa }$ are the
quantities defined in eqs. \ref{abd}, but with ${\bf Q}$ replaced by $%
\overrightarrow{{\bf \kappa }}=\widehat{{\bf z}}\;\omega $. Inserting these
results in eq. \ref{sgr} and summing over spins we get

\begin{eqnarray}
\frac{d\sigma _{\gamma }}{d\Omega }\left( \omega \right) &=&\frac{64\pi
e^{2}p}{a_{H}^{5}\omega {\rm v}_{+}}\frac{Z^{6}e^{2}}{\left[ \exp \left(
2\pi Ze^{2}/{\rm v}_{+}\right) -1\right] }\frac{1}{\left( \overrightarrow{%
{\bf \kappa }}{\bf -p}\right) ^{4}}  \nonumber \\
\times &&\Bigg\{\left( \varepsilon -1\right) \left[ \left( {\bf b}_{\kappa }%
{\bf -d}_{\kappa }\right) ^{2}-\frac{\left( {\bf p.}\widehat{{\bf e}}\right)
^{2}}{\left( \overrightarrow{{\bf \kappa }}{\bf -p}\right) ^{2}\left( \kappa
^{2}-p^{2}\right) }\right]  \nonumber \\
+ &&\left( \varepsilon +1\right) a_{\kappa }^{2}+2a_{\kappa }\left[ \frac{%
\left( {\bf p.}\widehat{{\bf e}}\right) ^{2}}{\left( \overrightarrow{{\bf %
\kappa }}{\bf -p}\right) ^{2}}-{\bf p}.({\bf b}_{\kappa }{\bf -d}_{\kappa })%
\right] \Bigg\}\;.  \label{srO}
\end{eqnarray}

We notice that the above equation can also be obtained from eq. \ref{st} in
the limit $q_{t}\rightarrow 0$ and ${\rm v}\longrightarrow c$. In this limit 
$\sigma _{\gamma ^{\ast }L}\longrightarrow 0$, and $\sigma _{\gamma ^{\ast
}LT}\longrightarrow 0$, and $\sigma _{\gamma ^{\ast }T}$ becomes the cross
section for the production of pairs, with capture, by real photons.

Integrating eq. \ref{srO} over the azimuthal angle we get (without any
further approximations!)

\begin{equation}
\frac{d\sigma _{\gamma }}{d\theta }\left( \omega \right) =4\pi ^{2}\frac{%
Z^{6}e^{2}}{\left[ \exp \left( 2\pi Ze^{2}/{\rm v}_{+}\right) -1\right] }%
\frac{e^{2}p^{3}\left( \varepsilon +2\right) }{a_{H}^{5}\omega ^{4}{\rm v}%
_{+}}\frac{\sin ^{3}\theta }{\left( \varepsilon -p\cos \theta \right) ^{4}}%
\left[ \varepsilon -p\cos \theta -\frac{2}{\omega (\varepsilon +2)}\right]
\label{srt}
\end{equation}

In the frame of reference of the nucleus, the angular distribution of the
positrons is forward peaked, along the incidence of the photon. The higher
the positron energy is, the more forward peaked the distribution becomes. In
a collider, a Lorentz transformation of these results to the laboratory
frame implies that all positrons are seen along the beam direction, within
an opening angle of order of $1/\gamma \ll 1$.

Integrating eq. \ref{srt} over $\theta $ we get

\begin{equation}
\sigma _{\gamma }=\frac{8\pi ^{2}e^{2}}{a_{H}^{5}\;{\rm v}_{+}}\frac{%
Z^{6}e^{2}}{\left[ \exp \left( 2\pi Ze^{2}/{\rm v}_{+}\right) -1\right] }%
\frac{p}{\omega ^{4}}\;\left[ \varepsilon ^{2}+\frac{2}{3}\varepsilon +\frac{%
4}{3}-\frac{\varepsilon +2}{p}\ln \left( \varepsilon +p\right) \right]
\label{sreal}
\end{equation}

The relevance of the Coulomb distortion corrections can be better understood
by using the high energy limit, $\varepsilon \gg 1.$ From eq. (\ref{sreal})
we get

\begin{equation}
\sigma _{\gamma }=\frac{8\pi ^{2}e^{2}}{a_{H}^{5}\;{\rm v}_{+}}\frac{%
Z^{6}e^{2}}{\left[ \exp \left( 2\pi Ze^{2}/{\rm v}_{+}\right) -1\right] }\;%
\frac{1}{\varepsilon }\;  \label{srealapp}
\end{equation}
If we had used plane-waves for the electron and hydrogenic wave function for
the positron in our calculation, without the corrections to order $Z\alpha $%
, the cross section for the pair production with capture by real photons
would be given by

\begin{equation}
\left[ \frac{d\sigma _{\gamma }}{d\theta }\left( \omega \right) \right]
_{PWA}=4\pi \frac{Z^{5}e^{2}p}{a_{H}^{5}\omega ^{4}}\frac{\sin \theta }{%
\left( \varepsilon -p\cos \theta \right) ^{4}}\;,
\end{equation}
the integral of which being

\begin{equation}
\left[ \sigma _{\gamma }\right] _{PWA}=\frac{8\pi Z^{5}e^{2}}{a_{H}^{5}}\;%
\frac{p}{\omega ^{4}}\;\left( 3\varepsilon ^{2}+p^{2}\right) \;.
\end{equation}
In the high energy limit:

\begin{equation}
\left[ \sigma _{\gamma }\right] _{PWA}=\frac{32\pi Z^{5}e^{2}}{3a_{H}^{5}}\;%
\frac{1}{\varepsilon }  \label{spwa}
\end{equation}

Using the approximation $Z\alpha \ll 1$ in eq. (\ref{srealapp}) we get

\begin{equation}
\sigma _{\gamma }=\frac{4\pi Z^{5}e^{2}}{a_{H}^{5}}\;\frac{1}{\varepsilon }
\label{spwa2}
\end{equation}
which is different than eq. (\ref{spwa}) by a factor 8/3. This is important
to show how the corrections of order $Z\alpha $ influence the result of the
calculation. Not only they modify the results by inclusion of the correct
normalization of the distorted positron wavefunction, $N_{+}$, but they also
yield terms to the matrix elements for photo-production of the same
magnitude as the terms of lowest order. The origin of the difference between
the two results are the small distances which enter in the calculation of
the matrix elements of eq. (\ref{Vfi}). These corrections are essential to
account for their effects properly, and are enough to account for a good
description of {pair }production with capture, as we will see in the next
section. It is worhtwhile to mention that this equation confirms the
findings of Sauter \cite{Sa31} \ for the annihilation of a positron with an
electron in the K-shell of an atom. Using the detailed balance theorem that
process can be related to the gamma production of electron-positron pairs
with electron capture and the above equation is reproduced.

All positron variables in the above equations are expressed in the frame of
reference of the ion where the electron is captured. Now the $({\bf Q}%
^{2}-\omega ^{2})^{2}$ denominator of the first term inside the integral \ref
{dsdedO} shows that $q_{t}\cong \omega /{\rm v}\gamma $ is the relevant
order of magnitude of $q_{t}$ in the integration. Thus, in all three virtual
photon cross sections, \ref{st}, \ref{sl}, and \ref{slt} we have $\left| 
{\bf Q}\right| \cong \omega /{\rm v}\cong \omega \gg \omega /{\rm v}\gamma $%
. The additional terms seem also to be independent from $\gamma $, what
would imply that no extra $\gamma $ factor arise in these expressions. Thus,
for large $\gamma $, $\sigma _{\gamma ^{\ast }T}\cong \gamma ^{2}\sigma
_{\gamma ^{\ast }L}\cong \gamma \sigma _{\gamma ^{\ast }LT}$. This is
confirmed by numerical calculations. Thus, for RHIC and LHC energies we can
safely use

\begin{equation}
\frac{d\sigma }{d\varepsilon d\Omega }=\frac{1}{\pi \omega }%
\;Z^{2}e^{2}\int_{0}^{\infty }d(q_{t}^{2})\;\frac{q_{t}^{2}}{({\bf Q}%
^{2}-\omega ^{2})^{2}}\;\Bigg[\frac{d\sigma _{\gamma ^{\ast }T}}{d\Omega }%
\left( {\bf Q},\omega \right) \Bigg]\;.  \label{sigrhic}
\end{equation}

The above discussion might seem convincing. But, it is still reasonable to
check this result within a solvable model. This can be achieved by using
plane-waves for the positron and hydrogenic waves for the electron, without
the correction terms to order $Z\alpha .$ In this case, we get

\begin{equation}
F({\bf Q})=4ie\sqrt{\pi }\left( Z/a_{H}\right) ^{5/2}\;\frac{\overline{{\rm v%
}}(\overrightarrow{\gamma }{\bf .K}){\rm u}}{\left( {\bf Q-p}\right) ^{2}}\;,
\end{equation}
which yields the cross section

\begin{equation}
\frac{d\sigma }{d\varepsilon d\Omega }=\frac{1}{\pi \omega }%
\;Z^{2}e^{2}\int_{0}^{\infty }d(q_{t}^{2})\;\frac{q_{t}^{2}}{({\bf Q}%
^{2}-\omega ^{2})^{2}}\;\Bigg[\frac{d\sigma _{\gamma ^{\ast }T}}{d\Omega }%
\left( {\bf Q},\omega \right) +\frac{d\sigma _{\gamma ^{\ast }L}}{d\Omega }%
\left( {\bf Q},\omega \right) \Bigg]\;,  \label{4.3}
\end{equation}
where the interference term is exactly zero, and

\begin{equation}
\frac{d\sigma _{\gamma ^{\ast }T}}{d\Omega }\left( {\bf Q},\omega \right) =%
\frac{32Z^{5}e^{2}p}{a_{H}^{5}}\frac{1}{\left( {\bf Q-p}\right) ^{8}}\;,\;\;%
{\rm and}\;\;\frac{d\sigma _{\gamma ^{\ast }L}}{d\Omega }\left( {\bf Q}%
,\omega \right) =\frac{32Z^{5}e^{2}p\omega ^{2}}{a_{H}^{5}q_{t}^{2}}\left( 
\frac{1}{\gamma ^{2}{\rm v}}\right) ^{2}\frac{1}{\left( {\bf Q-p}\right) ^{8}%
}\;.
\end{equation}

The integral over $\Omega $ in the expressions above can be done
analytically yielding

\begin{equation}
\left\{ 
\begin{tabular}{l}
$\sigma _{\gamma ^{\ast }T}\left( Q,\omega \right) $ \\ 
$\sigma _{\gamma ^{\ast }L}\left( Q,\omega \right) $%
\end{tabular}
\right\} =\frac{128\pi Z^{5}e^{2}p}{3a_{H}^{5}}\frac{\left( {\bf p}^{2}+3%
{\bf Q}^{2}\right) \left( 3{\bf p}^{2}+{\bf Q}^{2}\right) }{\left( {\bf Q}%
^{2}-{\bf p}^{2}\right) ^{6}}\left\{ 
\begin{tabular}{l}
$1$ \\ 
$\omega ^{2}/(q_{t}\gamma ^{2}{\rm v})^{2}$%
\end{tabular}
\right\} \;.
\end{equation}
Using this result, the integral over $q_{t}^{2}$ in (\ref{4.3}) can also be
performed analytically but resulting in too long expressions to be
transcribed here. We also note that the denominator $\left( {\bf Q-p}\right)
^{2}$ in eq. (\ref{FQ2}) implies that $\left| {\bf p}\right| \cong \left| 
{\bf Q}\right| \cong \omega /{\rm v}$ and that the largest contribution to
the integrals arise from the region with $\left( {\bf Q-p}\right) ^{2}\cong $
$\omega ^{2}/\gamma ^{2}$. Thus, in the plane wave approximation we really
expect that $\sigma _{\gamma ^{\ast }T}$ is by some power of $\gamma $\
larger than the $\sigma _{\gamma ^{\ast }L}$ cross section.

\section{ Total cross sections: comparison with the equivalent photon
approximation}

The energy spectrum of the positrons can be obtained by integrating \ref
{sigrhic} over angles. The calculation can be done analytically if one
neglects terms of order ${\cal O}\left( 1/\gamma \right) $. The expressions
obtained were checked against the numerical results and we found that for
RHIC energies the difference between the two results are only visible (of
order of 1\%) for positron energies $\varepsilon \ll 1$. For positron
energies $\varepsilon \gtrsim 1$ \ the differences are imperceptible. One
gets

\begin{equation}
\frac{d\sigma }{d\varepsilon }=\frac{32\pi p}{a_{H}^{5}\omega {\rm v}_{+}}\;%
\frac{Z^{8}e^{6}}{\left[ \exp \left( 2\pi Ze^{2}/{\rm v}_{+}\right) -1\right]
}\left\{ \frac{CD}{B^{6}}\left[ \ln \left( \frac{B}{A}\right) -\frac{197}{60}%
\right] +\frac{3}{5}\frac{D}{B^{5}}+\frac{1}{5}\frac{C}{B^{5}}+\frac{3}{20}%
\frac{1}{B^{4}}\right\} ,  \label{exact}
\end{equation}
where

\begin{equation}
A=\frac{\omega ^{2}}{\gamma ^{2}},\;\;B=\;\omega ^{2}-p^{2},\;\;C=3\omega
^{2}+p^{2},\;\;{\rm and}\;D=3p^{2}+\omega ^{2}.\;\;\;
\end{equation}

In ref. \cite{Be88} \ analytical formulas were also derived for the
differential and total cross section of pair production with K-shell
electron capture. The cross sections were obtained by using Sommerfeld-Maue
wave functions for the positron. Their result is

\begin{equation}
\frac{d\sigma }{d\epsilon }=\frac{16\pi p}{a_{H}^{5}\omega ^{5}{\rm v}_{+}}%
\frac{Z^{8}e^{6}}{\left[ \exp \left( 2\pi Ze^{2}/{\rm v}_{+}\right) -1\right]
}\left[ \frac{4}{3}+\frac{2\varepsilon }{3}+\varepsilon ^{2}-\frac{%
\varepsilon -2}{p}\ln \left( \varepsilon +p\right) \right] \ln \left( \frac{%
\gamma \delta }{\omega }\right) \;,  \label{dsigde}
\end{equation}
where $\delta =0.68108...$ is related to the Euler's number.

It is also worthwhile to compare the calculations with the equivalent photon
approximation. The function

\begin{equation}
n\left( \gamma ,Q\right) =\frac{1}{\pi }\;Z^{2}e^{2}\frac{\;q_{t}^{2}}{\left[
q_{t}^{2}+\left( \omega /\gamma {\rm v}\right) ^{2}\right] ^{2}}  \label{epn}
\end{equation}
is called by ``equivalent photon number'', and is strongly peaked at $%
q_{t}\simeq \omega /\gamma {\rm v}$, which is very small for $\gamma \gg 1$.
The equivalent photon approximation states that

\begin{equation}
\frac{d\sigma }{d\varepsilon }\simeq \int_{0}^{q_{t}^{\max }}d\left(
q_{t}^{2}\right) \;\frac{1}{\omega }\;n\left( \gamma ,Q\right) \;\sigma
_{\gamma ^{\ast }T}\left( q_{t}=0,\omega \right) =\frac{1}{\omega }\;\sigma
_{\gamma }\left( \omega \right) \;\int_{0}^{q_{t}^{\max }}d\left(
q_{t}^{2}\right) \;n\left( \gamma ,Q\right)  \label{epn2}
\end{equation}
where in the last equality one approximates $\sigma _{\gamma ^{\ast
}T}\left( q_{t}=0,\omega \right) \simeq \sigma _{\gamma }\left( \omega
\right) .$ The problem with the approximation is that the integral in eq. 
\ref{epn2} diverges logarithmically. The approximation is only valid if we
include a cutoff parameter $q_{t}^{\max }$, determined by the value of $%
q_{t} $ at which $\sigma _{\gamma ^{\ast }T}\left( Q,\omega \right) $ drops
to zero. For RHIC and LHC energies, we verified that the transverse virtual
photon cross section drops to zero at $q_{t}\simeq \varepsilon .$ Another
hint for obtaining the appropriate cutoff value of $q_{t}$ is given in
figure 2 where we plot the energy spectrum of the positron, $d\sigma
/d\varepsilon $, for RHIC energies. We observe that the energy spectrum
peaks at $\varepsilon \simeq 2-3$. We thus conclude that an appropriate
value of the cutoff parameter is $q_{t}^{\max }\sim 1$. Inserting this value
in eq. \ref{epn2} we get the same result as \ref{dsigde}, except that in the
last logarithm of that formula $\delta $ is replaced by $0.6065...$, an
irrelevant change, in view of the large value of $\gamma $.

In figure 2 we compare our result, eq. \ref{exact} (solid line) with the
approximation \ref{dsigde} (dashed line). We observe a good agreement,
except at the low positron energies. This is an important result. The
Sommerfeld-Maue wave functions (see \cite{Be54} for a complete discussion
about these wavefunctions) reproduce the exact Coulomb-Dirac wavefunction in
the region relevant for the pair-production process. They account for the
distortion effects in all orders. The result of figure 2 shows that, for
pair production with K-shell capture, the approximation \ref{phip}, with \ref
{phipq}, leads to very close agreement with that using Sommerfeld-Maue
wavefunctions. That is, the most important part of the distortion is already
taken into account by expanding the wavefunction to first-order in $Z\alpha $
and by using the correct normalization for the distorted wave \ref{NN}. We
also note from figure 2 that the two results agree at large positron
energies, $\varepsilon \gg 1.$ That is expected, since distortion effects
are not important at these energies and the outgoing positron is well
described by plane-waves.

The total cross section can be obtained from eqs \ref{exact} and \ref{dsigde}
by an integration over the positron energy. The integrations are performed
numerically. In ref. \cite{Be88} an approximate analytical result was
obtained. Their result is

\begin{equation}
\sigma =\frac{33\pi }{10a_{H}^{5}}Z^{8}e^{6}\frac{1}{\exp \{2\pi Ze^{2}\}-1}%
\left[ \ln \left( \gamma \delta /2\right) -5/3\right] .  \label{bbb}
\end{equation}

The term $\left[ {}\right] ^{-1}$ is the main efffect of the distortion of
the positron wavefunction. It arises through the normalization of the
continuum wavefunctions which accounts for the reduction of the magnitude of
the positron wavefunction near the nucleus where the electron is localized
(bound). Thus, the greater the $Z$, the less these wavefunctions overlap.

In figure 3 we compare the three results for RHIC energies, as a function of
the atomic charge $Z$. The numerical integration of eqs. \ref{exact} and \ref
{dsigde} yields essentially the same result, differing at most within 2\%
from each other. They are displayed by the solid curve. The dashed curve is
the result of \ref{bbb}. It agrees with the integrations of eqs. \ref{exact}
and \ref{dsigde} for low Z ($Z\lesssim 15$) but disagrees by \ a factor 5
for large Z ($Z\sim 80-90$). The reason for this discrepancy is that eq. \ref
{bbb} was deduced in ref. \cite{Be88} by probably neglecting the strong
energy dependence (for large Z) of the exponential in the denominator of \ref
{exact} which is a function of the positron velocity, ${\rm v}_{+}$.

We note that the cross section \ref{exact} scales approximately with $Z^{7}$
for small values of $Z$, as also implied by eq. \ref{bbb}. But, for large $Z$%
's there is not such a simple dependence. In our case, it is directly
obtained from integration of eq. \ref{exact}. The charge dependence of the
capture cross sections have also been studied in refs. \cite{As94,Ag97} with
slightly different results. For asymmetric systems, e.g., a projectile of
charge $Z_{P}$ incident on a target of charge $Z_{T\text{ }}$, the equation 
\ref{exact} should be changed by replacing $Z^{8}$ in the numerator by $%
Z_{P}^{2}Z_{T}^{6}$ and $Z$ in the exponential term of the denominator by $%
Z_{T}$.

Equation \ref{exact} (and also \ref{bbb}) predicts a dependence of the cross
section in the form $\sigma =A\ln \gamma _{c}+B$, where $A$ and $B$ are
coefficients depending on the system and bombarding energy \cite{Be88}. This
dependence was used in the analysis of the experiment in ref. \cite{Kr98}.
In recent calculations, attention was given to the correct teatment of the
distortion effects in the positron wavefunction \cite{He00}. In figure 4 we
show the recent experimental data of ref. \cite{Kr98} compared to eq. \ref
{bbb} (dashed line) and to eq. \ref{exact}. The data was obtained for $%
Pb^{82+}$ at 33 TeV impinging on several targets. The comparison with the
experimental data is not fair since atomic screening was not taken into
account here. When screening is present the cross sections will always be
smaller at least by a factor 2-4 \cite{Be88}. 

Using eq. \ref{exact} for $Au+Au$ beams at RHIC (100 GeV/nucleon) and $Pb+Pb$
beams at LHC (3 TeV/nucleon), we get $\sigma =229$ b and $\sigma =562$ b,
respectively. Other results can be easily obtained from a simple integration
of eq. \ref{exact}. In ref. \cite{BRW93} a numerical calculation was done
making use of full Dirac wavefunctions, yielding 114 b for $Au+Au$ beams at
RHIC. This is about 2 times larger than that obtaining using eq. \ref{bbb}
from ref. \cite{Be88} and about 2 times smaller than our eq. \ref{exact}. We
assume that the reason for these diferences are that our wavefunctions are
correct only to first order in $Z\alpha $. It might be that the inclusion of
higher-order terms will change our results. However, the nice feature of our
approach is that it provides an analytical description of the process and a
direct connection with the photo-production cross sections.

\section{Conclusions}

We have calculated the cross sections for pair production with K-shell
capture in peripheral collisions of relativistic heavy ions. The cross
section can be separated into longitudinal and transverse components, with
respect to the beam direction. For high energies, $\gamma \gg 1$, this
corresponds to the usual meaning of longitudinal and transverse components
of the photon. A very transparent and simple formulation is obtained using
the lowest order corrections to the positron and electron wavefunctions. At
ultra-relativistic energies, corresponding to the energies of relativistic
heavy ion colliders (RHIC and LHC), the contribution of the longitudinal
virtual photons to the total cross section vanishes. The remaining terms of
the cross section can be factorized in terms of a virtual photon spectra and
the cross section induced by real photons. However, the factorization
depends on a cutoff parameter, which is not well defined. But, for $\gamma
\gg 1$ a momentum cutoff parameter $q_{t}\sim 1$ reproduces extremely well
the exact value of the cross section.

We note that the electron can be captured in any atomic orbit around one of
the ions, the capture to the K-shell being the largest contribution. In ref. 
\cite{Be88} it was shown that the contribution of capture to all other
shells will contribute by an increase of 20\% of the presently calculated
cross section. Depending on the relevance of this process to peripheral
collisions with relativistic heavy ions and on the accuracy attained by the
experiments, more theoretical studies on the capture to higher orbits should
be performed. \ At the present stage, it is important to notice that eq. \ref
{bbb}, from ref. \cite{Be88}, is not a good approximation for large Z ions.
The calculations developed in the present article were important to pinpoint
the origin of this discrepancy and the validity of the equivalent photon
approximation for $\gamma \gg 1$. It would be nice to show in a future study
if the distortion corrections of higher order in $Z\alpha $ are indeed
necessary to account for. 

\bigskip

{\Large {\bf Acknowledgments}}

(*) John Simon Guggenheim Fellow.

This work was partially supported by the Brazilian funding agencies CNPq,
CAPES, FUJB and MCT/FINEP/CNPQ(PRONEX) under contract No. 41.96.0886.00.
E-mails: bertu@if.ufrj.br, dolci@if.ufrj.br.

\section{Figure captions}

\begin{enumerate}
\item  {} Diagram for pair production with capture in relativistic heavy ion
colliders.

\item  {} Energy spectrum of the emitted positron in pair production with
capture for 100 GeV/nucleon $Pb+Pb$ collisions at RHIC in the frame of
reference of the nucleus where the electron is captured. The differential
cross section is multiplied by the electron rest mass. The cross section is
given in barns. The solid line is the exact calculation, while the dashed
line is the analytical approximation, eq. \ref{dsigde}.

\item  {} Total cross sections for pair production at RHIC as a function of
the nuclear charge of the ions. The solid line is the exact calculation, and
the dashed line is obtained from eq. \ref{bbb} given in ref. \cite{Be88}.

\item  Cross sections for pair production with electron capture for $Pb^{82+}
$ at 33 TeV impinging on several targets with charge $Z$. The recent
experimental data of ref. \cite{Kr98} is compared to eq. \ref{bbb} (dashed
line) and to eq. \ref{exact} (solid line). 
\end{enumerate}

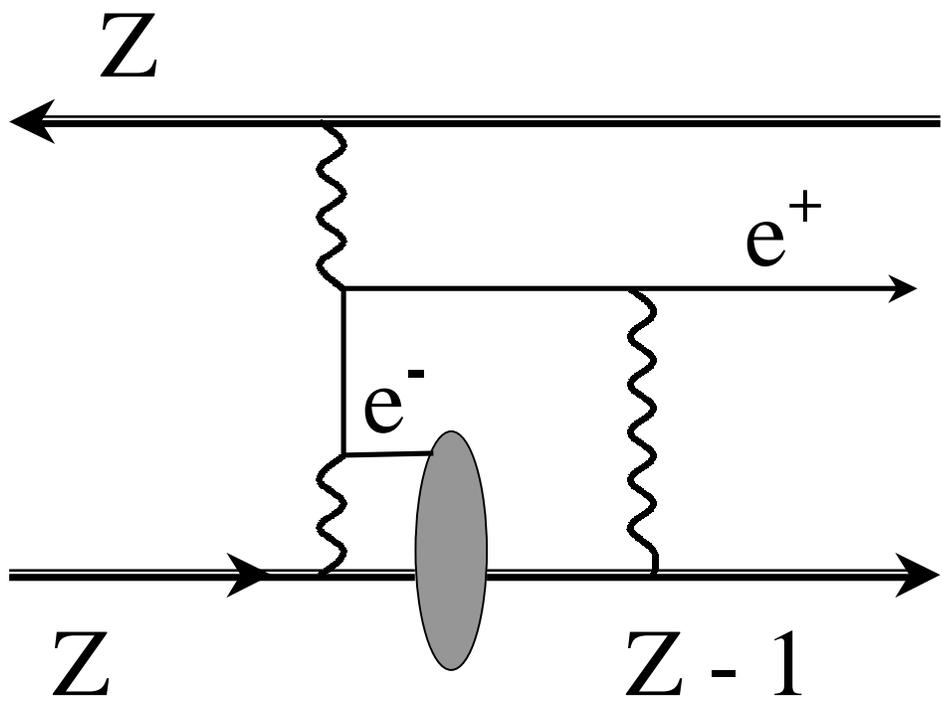
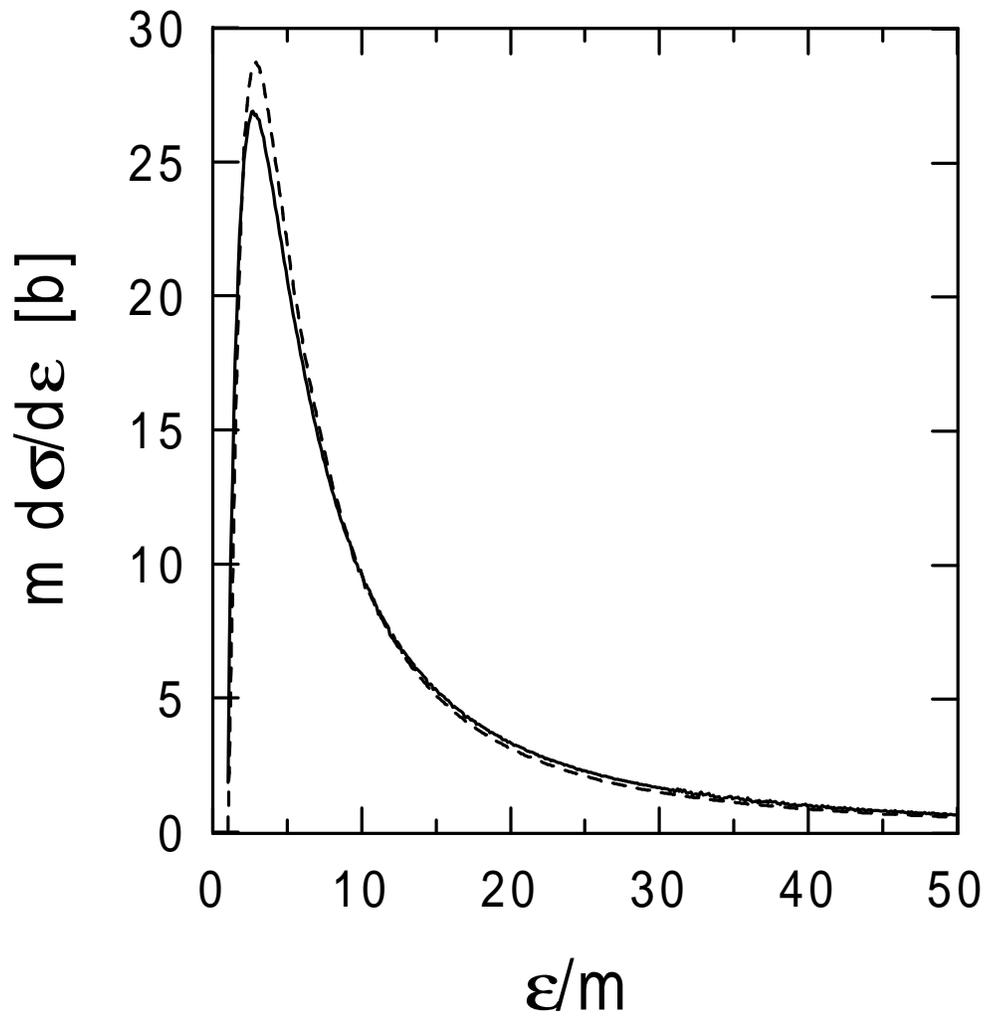
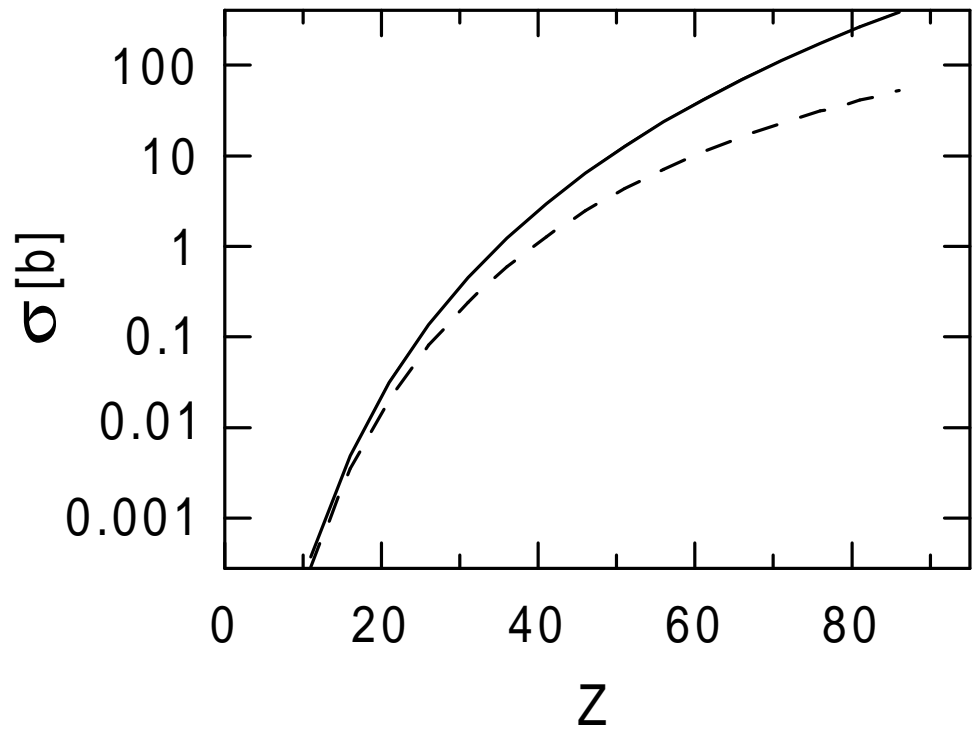
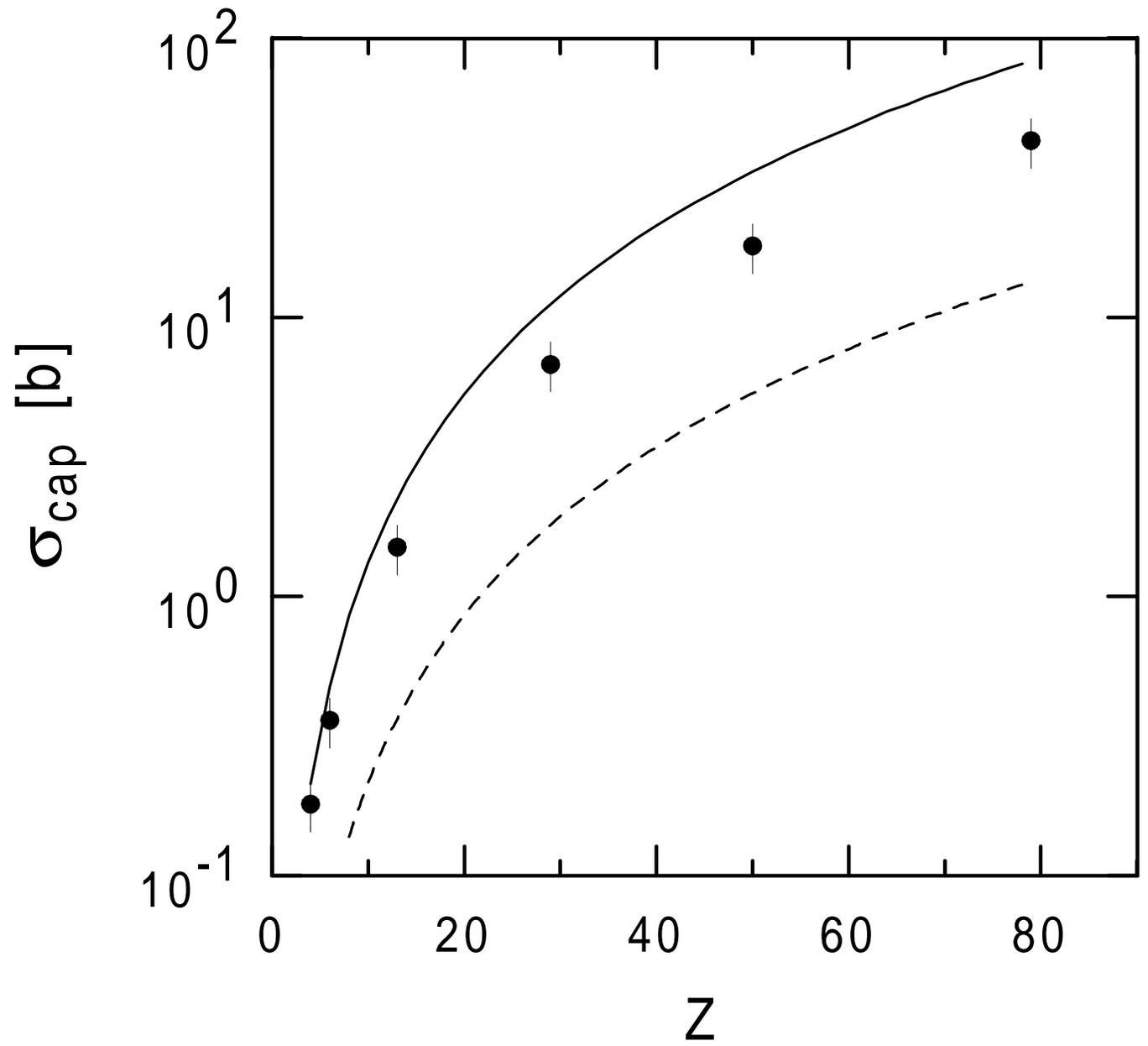
\end{document}